# DESIGN OF APRS MODEM USING IC TCM3105 AND ATMEGA2560 MICROCONTROLLER

Rommy Hartono
Satellite Technology Center
National Institute of Aeronautics and Space (LAPAN)
Bogor, Indonesia
rommy.hartono@lapan.go.id

Isma Choiriyah
Satellite Technology Center
National Institute of Aeronautics and Space (LAPAN)
Bogor, Indonesia
isma.choiriyah@lapan.go.id

Wahyudi Hasbi
Satellite Technology Center
National Institute of Aeronautics and Space (LAPAN)
Bogor, Indonesia
wahyudi.hasbi@lapan.go.id

Rakhmad Yatim
Satellite Technology Center
National Institute of Aeronautics and Space (LAPAN)
Bogor, Indonesia
rakhmad.yatim@lapan.go.id

*Abstract*—**APRS technology is still exists and still developing among amateur radio. However, in Indonesia there are still not many people who use and utilize APRS. The high price of APRS modem and difficulties of getting APRS modems is one of the factors affecting the public interest in APRS. In Indonesia, no one has developed APRS modem, so that our paper will be discuss about design and develop of APRS modem using cheap and easy components FSK modem IC TCM3105 and ATmega2560 as microcontroller. The implementation of APRS standard protocol on modem that has been made use C programming language on Arduino IDE and the schematic design and printed circuit board layout use Proteus 8. IC TCM3105 selection is based on the specification of eligible components of APRS device able to set the baud rate of 1200 bps with FSK modulation technique. The ATmega2560 microcontroller is chosen as the information signal encoding processor and calculate the CRC-16-X25 FCS (Frame Check Sequence) for the information packet according to the standard AX.25 UI Frame APRS protocol. AX.25 UI Frame APRS protocol consists of flag, destination address, source address, digipeater address, control field, protocol ID, information field and FCS. The result of this paper is APRS modem device that has been made can send the APRS information packet very well. This is proved by APRS packet that has been sent can be received and repeated by ground station of Pusteksat LAPAN also connected on APRS international network in aprs.fi. In addition, the information that has been sent from APRS modem can be decoded correctly by APRS software decoder such as Soundmodem, AFSK1200, dan AX.25-SCS.**

## I. INTRODUCTION

Automatic Packet Reporting System (APRS) is a radio packet application to transmit information about position, short text message and data telemetry. The APRS has been registered and developed by a member of amateur radio called Bob Bruninga, callsign WB4APR [1]. APRS is a basic system of digital communications and information for real time communication in the local area. APRS is similar technology with Global Positioning System (GPS), a technology to display on map a local area through the satellite. But, it is different with GPS, APRS is made by individual or group or institution and it is used for themselves but it does not rule out that other people can see the visualisation this APRS. An APRS station can change data from sensors or position that is shown by GPS become radio packet format (AX.25 UI Frame) then will be transmitted through radio wave with speed 1200 bps for Very High Frequency (VHF) band or 300 bps for High Frequency (HF) band [2].

APRS also can be used as an alternative communication for disaster areas. APRS can increase the effectiveness to help others[3]. It is possible to convey information quickly, and accurately, so the assistance can quickly come and can minimize the number of victims who get worse[4].

Basically, APRS needs devices to build system include power supply, a number of sensors and transducers, radio transceiver, APRS transmitter and modem [5][13]. Fig. 1 shows about the block diagram of transmission and reception system in APRS. In general, the source information will be encoded by the APRS modem before will be transmitted by the transmitter through the radio waves. Then in receiver, the received signal will be encoded and become the information as sent by the transmitter.

APRS modem is one of the most important parts of the process of sending and receiving APRS packet, which the modem works as digital signal converters into analog signal before information is transmitted through radio wave and convert analog signals into digital signals when receiving APRS packet information. Currently, the availability of APRS modem in Indonesia still does not yet exist, because the APRS modem must be brought directly from abroad.

In this paper, we will discuss about the design of APRS modem using different component with existing APRS modems. This APRS modem is built using the IC TCM3105 modem and ATMega2560 microcontroller. IC TCM3105 has function to adjust the baud rate and modulate the signal before the information is sent. The APRS uses AX.25 packets with 1200 baud audio frequency shift keying (AFSK) modulation [6]. Then the ATMega2560 microcontroller as a processor for the information signal coding process as well as the CRC-16-X25 Frame Check Sequence (FCS) counter so that the information packet complies with the APRS protocol standard, AX.25 UI Frame.

The limitations of the discussion of this study is only on the APRS packet transmission system not yet reaching on the reception system. The purpose of designing this APRS modem is the creation of a new APRS modem device that has lower price than the existing APRS modem device.

The rest of this paper is organized as follows. First, the methodology of this research will be determined briefly. It presents the step that will be done in this research from the literature study until testing and analysis. Each step explain the result in result and discussion section. The system design is implemented by using C programming language on Arduino IDE and Proteus 8 for the printed circuit board. Then, the testing is done to the system. Finally, the result is analyzed and the system works successfully as the system design.

## II. METHODOLOGY

The methodology of this research can be seen in fig. 2. Fig.2 shows the steps of the author to do this research, such as:

1) Literature study: study and collect the references about standard protocol used in APRS;
2) Observation: observe the audio signal from APRS packet as a modulated information using Frequency Shift Keying (FSK) modulation.
3) System design: design the low cost APRS device using IC TCM3105 (FSK Modulator-Demodulator) and microcontroller ATMega2560 (information processor, encoding).

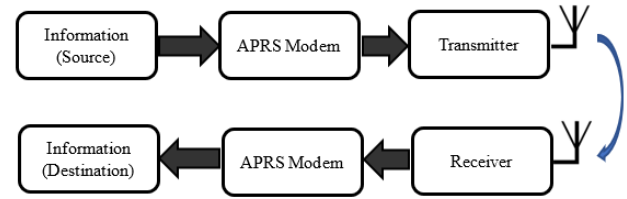

Figure 1. Block Diagram of Transmission and Reception System in APRS

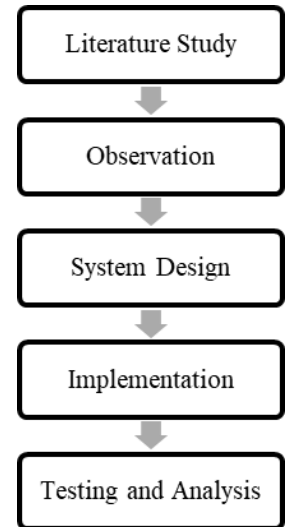

Figure 2. Methodology

4) Implementation: shows the process of making the hardware. The process is start from schematic circuit and printed Circuit Board (PCB) based on design results use Proteus 8.
5) Testing and analysis: testing of APRS packet transmission uses TCM3105 modem with an instrumentation aid and an existing APRS modem.

## III. RESULT AND DISCUSSION

### A. Literature Study

Literature study has function as the beginning of the author to learn about radio communication system especially APRS. APRS is packet communication protocol to disseminate data directly to everyone on the network in real time which allows the feature of a combination radio package with the Global Positioning System (GPS), so that, the amateur radio can automatically show the position of radio station and another object in map [7].

Based on that, before designing an APRS modem, the most important thing is to understand the standard protocol that has to be used. At data link layer, APRS uses AX.25 protocol, as defined in Amateur Packet-Radio Link Layer Protocol with especially using Unnumbered Information (UI) [8][14]. The format of AX.25 UI-Frame packet can be seen in Table I.

TABLE I. AX.25 UI-FRAME PACKET FORMAT

| AX.25 UI-FRAME FORMAT | | | | | | | | |
|---|---|---|---|---|---|---|---|---|
| Flag | Destination Address | Source Address | Digipeater Addresses (0-8) | Control Field | Protocol ID | Information Field | FCS | Flag |
| 1 | 7 | 7 | 0-56 | 1 | 1 | 1-256 | 2 | 1 *(byte)* |

Table I shows the format of AX.25 UI-Frame packet for APRS. It consists of several parts such as:

1) **Flag**

The flag field is bit sequence 0x7E that is inserted at each end of the frame and separates each frame. Two frames may share one flag, which would denote the end of the first frame and the start of the next frame [9].

2) **Address**

An address contains 7 bytes that consists of 6 bytes of callsign and 1 byte of Secondary Station Identifier (SSID). Callsign consist 6 upper-case letters or numbers ASCII characters only. SSID is 4-bit integer that identifies multiple stations uniquely using same callsign [9]. There are several kinds of addresses in AX.25 UI-Frame format, that is:

a. **Destination Address**
The destination address field contain an APRS destination callsign and SSID [8].

b. **Source Address**
The source address field contains the callsign and SSID of the transmitting station. In some cases, if the SSID is non-zero, the SSID may specify an APRS display Symbol Code.

c. **Digipeater Addresses**
The digipeater addresses field contain the callsign of digipeater. This field can be filled up to 8 digipeater addresses [8].

3) **Control Field**

The control field identifies the type of frame being passed and controls several attributes of the layer 2 connection. APRS uses AX.25 UI-frame protocol so that this field is set to be 0x03.

4) **Protocol ID**

The Protocol ID field is set to be 0xF0 (no layer 3 protocol).

5) **Information Field**

The information field contains data of APRS such as short text message, sensors data, announcement or other data. The first character of this field is the APRS Data Type Identifier that specifies the nature of the data that follows. The maximum size of information field is 256 bytes.

6) **Frame Check Sequence (FCS)**

FCS field is a 16-bit number that calculated by the transmitter or receiver station. This is will ensure the integrity of the received frame or to make sure that the frame is not corrupted when transmitted. FCS is such a CRC that calculated by polynomial of $x^{16} + x^{12} + x^5 + 1$ (CRC-CITT/X-25).

### B. Observation

Based on Bell 202 standard for FSK modulation in baud rate 1200bps, mark frequency is 1200Hz as representation of logic '1' and space frequency is 2200 Hz as representation of logic '0'. The observations have function to authenticate the compatibility mark-space frequency of the TCM3105 FSK modulated audio signal with audio signals from OpenTracker-USB APRS devices which are processed by Matlab software.

Based on the result in Fig. 3, it shows that the information bit which is sent via TXD pin of IC TCM3105 produce modulated audio that matches with the mark and space frequency of the APRS audio signal which is analyzed by Matlab. But, the logic of the signal is opposite to the bit of information data that is sent, so that in the process of sending data bits from the microcontroller necessary needs to invert the logic bits. In addition, the process of sending data bit from or to the TXD pin of TCM3105 must be done in raw serial bits so that it cannot be used by the serial UART protocol because in the serial UART there are start bits and stop bits in it [10].

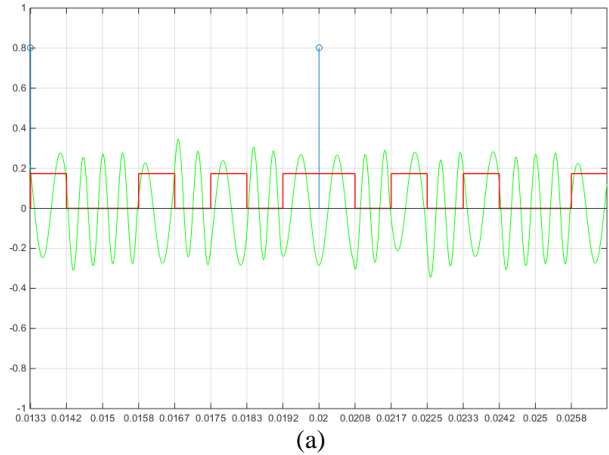

(a)

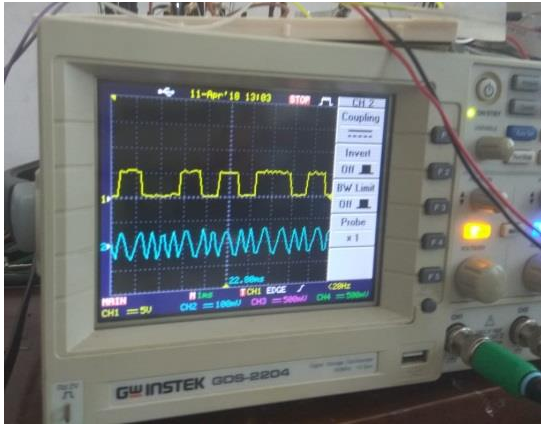

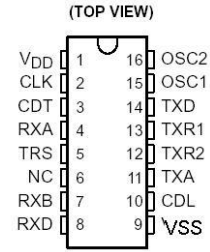

Figure 4. IC TCM3105 Pins

Figure 3. Observation of mark and space frequency: (a) FSK Audio Signal in Matlab; (b) Output from FSK Audio Signal of IC TCM3105

## C. System Design

The whole system consists of hardware and software. The hardware is IC TCM3105 as a FSK modulator-demodulator and an ATMega2560 microcontroller as a data processor and coding algorithm. TCM3105 is a single-chip asynchronous FSK modem that uses CMOS technology [11]. It has 16 pins as shown in Fig. 4.

The data bit rate used are 1200 bps according to the bit rate specifications of commonly used APRS devices. The selection of bit rate 1200 bps is done by configuring the TXR1 and TXR2 pins that is connected to VSS (Ground) while the TRS pin is given an invert value from the CLK pin with a NOT logic gate in accordance with the "operating mode" which is listed on the IC TCM3105 datasheet. Completely, the TCM3105 modem circuit that we built is as shown as on Fig. 5.

The differences between TCM3105 modem circuit that we developed and the existing APRS modem using TCM3105 are on RXB (RX Bias) pin and CDL (Carrier Detect Level) pin. The voltage value of RXB pin is 2.7 V (2.3 V – 3.1 V) and the reference voltage of CDL pin is 3.3 V (2.8 V – 3.9 V) [11].

The design of algorithm for program on ATMega2560 microcontroller used in sending APRS packet information in accordance with the AX.25 UI-Frame protocol standard is as follows:

1. Define the Destination Address+SSID, Source Address+SSID, Path (if needed), Control Field, Protocol ID and the Information Field.
2. Do the bit shifting, 1 bit to the left (<<1) for each byte of call sign from Destination, Source and Path.
3. Calculate the FCS (CRC-CCIT/X-25), process the CRC for each data byte start from Destination Address until Information Field using $x^{16} + x^{12} + x^5 + 1$ polynomial, input value of XOR is 0xFFFF, and output value of XOR is 0xFFFF.

4. Add the FCS to the packet with exchanging the order between the position of first octet FCS byte and the second octet FCS byte.
5. Exchange the bit sequence (bit swapping) between MSB and LSB of each byte of data
6. Insert the bit stuffing according to HDLC conditions (High-Level Data Link Control)
7. Add the flag which is 0x7E at the beginning and end of the package
8. Perform NRZ-I (Non-Returned to Zero Inverted) encoding on all data bytes
9. Send the data bit in raw bits serially to the TXD pin of IC TCM3105.

## D. Implementation

The microcontroller as an information processor that we use is ATMega2560 with a minimum system in the form of Arduino Mega board with C programming language on Arduino IDE. While, the TCM3105 modem circuit is implemented in the form of modules that can be directly paired on the Arduino Mega board. In more detail, the hardware implementation based on the design results is shown on Fig.6.

Fig. 6 shows the design PCB of TCM3105 Modem Circuit and its 3D view using Proteus8. The modem circuit that we built also has been equipped with a series of Push-To-Talk (PTT) automatically when it gets a trigger from Arduino Mega2560 pin 2. The audio output terminal is connected to a 10k resistor which is driven by a NPN transistor, so that when the base of the NPN transistor gets a current flow, the PTT of the Handy Talky (HT) radio will be active because it is also connected to the TCM3105 audio modem terminal. The hardware implementation of APRS modem can be seen on Fig. 7. The result of the implementation of a module circuit that has been integrated with Arduino Mega2560 only requires a 5 V DC supply voltage as shown.

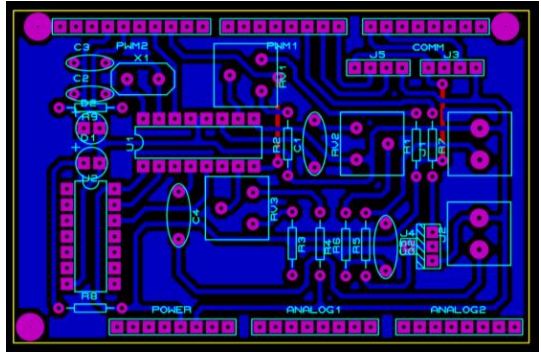

(a)

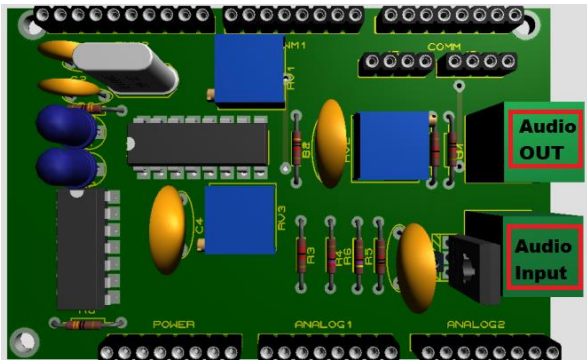

(b)

Figure 6. (a) Design PCB of TCM3105 Modem Circuit on Proteus 8;
(b) 3D View Design on Proteus 8

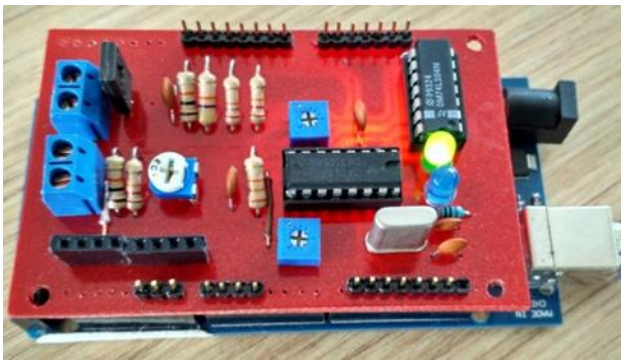

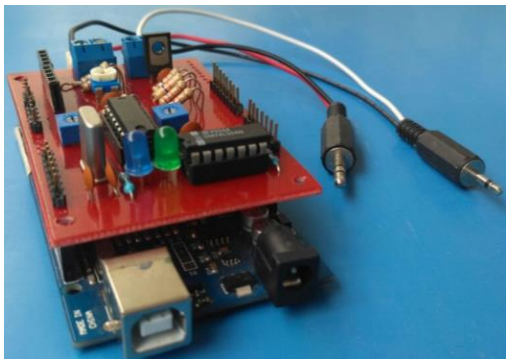

Figure 7. Hardware of APRS Modem

*E. Testing and Analysis*

The test is carried out using several devices, such as hardware like HT radio as a medium for transmission via radio waves and Software Define Radio (SDR) dongle as the recipient. The testing software used is HDSDR as a radio interface and the APRS Audio signal decoder uses Soundmodem, AFSK1200, and AX.25-SCS[12]. The testing scheme used is shown on Fig. 8. Fig. 8(a) shows the hardware setup for testing and Fig. 8(b) shows the scheme for testing process.

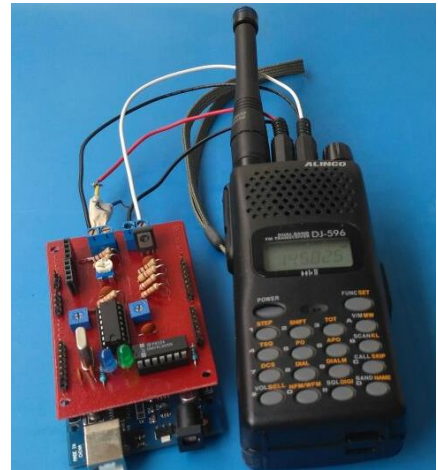

(a)

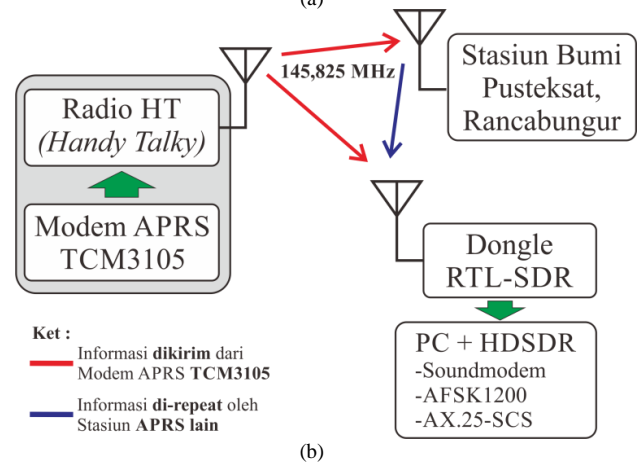

(b)

Figure 8. System testing: (a) Modem Setup; (b) Testing Scheme

The first testing scheme carried out was by sending the APRS packet using the TCM3105 modem via HT Alinco DJ-596 Radio at a frequency of 145.825 MHz and the receiving part used the RTL-SDR dongle and HDSDR software with VB-Audio Virtual Cable as the audio pipeline[15]. Then, it used 3 different APRS decoder software as comparison material for acceptance signal as seen on Fig. 9. Fig. 9 shows the comparison of APRS packet acceptance test using 3 different APRS decoder software, Soundmodem, AFSK1200 and AX.25-SCS.

That 3 software shows the same result for the received signal.

After testing the packet transmission received by our own radio station using RTL-SDR dongle, the next is testing the APRS packet transmission through a station that has function as digipeater. We do the test in around Satellite Technology Center (Pusteksat) LAPAN, Rancabungur West Java. Pusteksat APRS station has a callsign YF1ZQA and LAPAN-A2 satellite that has APRS mission has callsign YBSAT. So, we conducted a test to send the APRS packet by adding the YBSAT callsign as the path address and also WIDE2-2 so that the APRS package will be repeated. The packet details that we send can be seen in Table 2. We use APTCM0 as destination address. The destination address on the modem is the software version of the APRS modem used. Because there is no APRS modem that uses TCM3105 in the market and is registered in the "TOCALL" APRS, so APTCM0 is used as a prototype for the APRS TCM3105 modem.

When we test the APRS packet transmission via TCM3105 modem, we also monitor by using RTL-SDR dongle to know if the APRS packet was received by another APRS station successfully. Based on the test results on the Fig. 10, the APRS packet that we sent via the TCM3105 modem was successfully received and also repeated by the YF1ZQA APRS station (Pusteksat, Rancabungur). Moreover, we also can check the result by accessing the web aprs.fi and search the callsign. In Fig. 11 we know that the APRS packet that we sent via TCM3105 modem was also successfully transmitted to the aprs.fi through Internet Gateway (I-Gate) from YF1ZQA APRS station (Pusteksat, Rancabungur).

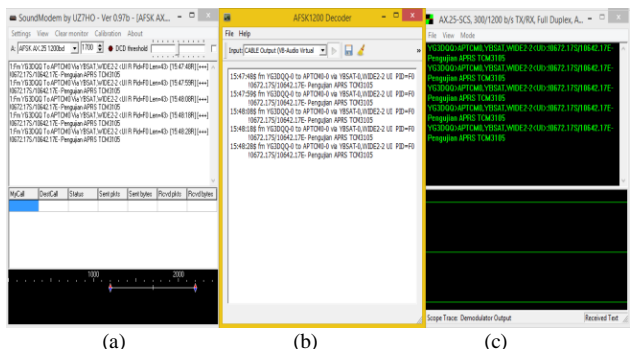

Figure 9. APRS Packet Acceptance Test: (a) Soundmodem; (b) AFSK1200; (c) AX.25-SCS

TABLE 2. DETAILS OF THE APRS TEST PACKET

| Destination Address | Source Address | Path Address | Information |
|---|---|---|---|
| APTCM0 | YG3DQQ | YBSAT, WIDE2-2 | Pengujian APRS TCM3105 |

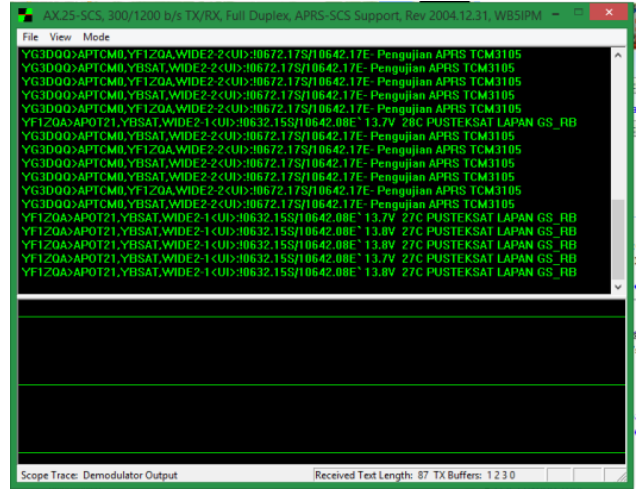

Figure 10. System Test Result

## IV. CONCLUSION AND FUTURE WORKS

Based on the results of the discussion and analysis, the following conclusions are taken:

- The Making of APRS modems at low cost can be realized by referring to the existing APRS modem devices

- The APRS modem device that has been made has been tested and the result is that the APRS information packet can be sent properly. This is proven by the result that the APRS packet was received and repeated by the LAPAN Pusteksat ground station so that it entered the international APRS network at aprs.fi.

- A way to find out that the information packet is appropriate with the APRS protocol is by decoding the packet information using APRS decoder software such as Soundmodem, AFSK1200, and AX.25-SCS. And the packet that is sent by TCM3105 modem has been successfully decoded by that 3-software decoder.

- This APRS modem has a separate processing function between the analog signal modulators using FSK modulation techniques and the microcontrollers as data information processing so that the modem is more flexible to be developed.

The future works for this research is there will be a comparison between APRS TCM3105 modem of this research and the existing APRS modem devices such as Kantronics, APRS Voyager, etc. That comparison will indicate the reliability of this APRS TCM3105 modem device than the existing.

## ACKNOWLEDGMENT

The authors would like to thank Mr. Mujtahid as Director of LAPAN Satellite Technology Center and

Group Leader Operation of LAPAN -A2 and LAPAN-A3 Satellite Mr. Patria Rachman Hakim and also thank you to LAPAN satellite operators especially Ground Support for their support and assistance so that this work can be well completed.

## REFERENCES

[1] Bicket, Tom, "Automatic Position Reporting System (APRS)",2010,Radio Amateur Association.

[2] Dear, Varuliantor,"The potential for using the APRS system to disseminate information on space weather conditions. Potensi pemanfaatan sistem aprs untuk sarana penyebaran informasi kondisi cuaca antariksa",2010, Berita Dirgantara, LAPAN. 11 (3):72-79.

[3] Sri, Aji K.,"Dual mitigation system : database system combination of EWS and APRS for disaster management (case study : Malang southern coast)",2015,CITIES 2015,Procedia Social and Behavioral Sciences

[4] A. Goeritno, "APRS implementation for monitoring and measurement data packages. Implementasi APRS untuk paket data pemantauan dan pengukuran", 2014, SETRUM Vol. 3 No 2 Desember 2014

[5] Adisoemarta, S.,"APRS and its application. APRS dan aplikasinya", 2008, Proceeding SIPTEKGAN XII 2008, pp. 749-757.

[6] S. Toledo, "A high-performance sound-card AX.25 modem", 2012, QEX-July Agustus 2012.

[7] Karn, P.R., Price, H.E., and Diersing, R.J.,1985, Packet radio in the amateur service, IEEE J. Select. Areas Commun., 3(3), 431–439.

[8] Automatic position reporting system (APRS) protocol reference Protocol Version 1.0, Tucson Amateur Packet Radio Corp, 2000.

[9] George, Florian. Telemetry and Telecommand Transfer Frames Format, SwissCube, Netherland, 2007.

[10] Hansen, John. PIC-et Radio: How to Send AX.25 UI Frames Using Inexpensive PIC Microprocessors, ARRL AND TAPR DIGITAL COMMUNICATIONS CONFERENCE, American Radio Relay Leage, Newington, 1998.

[11] Texas Instrument, 1994, TCM3105 FSK MODEM, in www.netti.fi/~ryydis/tcm3105.pdf, accessed on March 2018.

[12] "Monitoring APRS with the RTL-SDR" in www.rtl-sdr.com, accessed on August 2018

[13] Dąbrowski, K.,1994, Digital amateur communications. PWN, Warsaw. In Polish

[14] Zieliński, B.,2009, An analytical model of TNC controller, Theoretical and Applied Informatics, 21(1), 7–22

[15] Author of RTL-SDR.com, 2012, A guide to RTL-SDR and cheap software defined radio, The hobbyist's guide to RTL-SDR

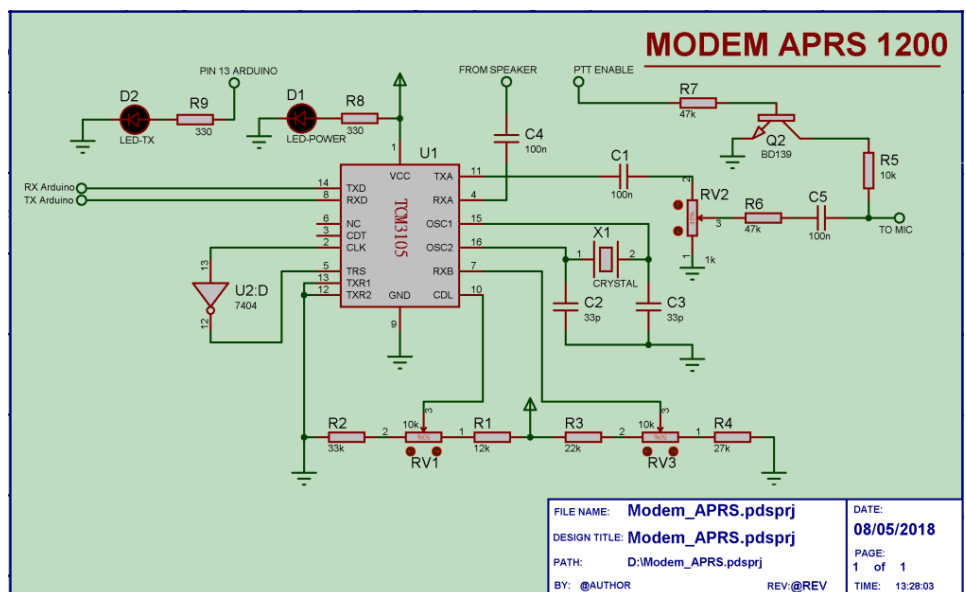

Figure 5. Schematic of APRS Modem Circuit

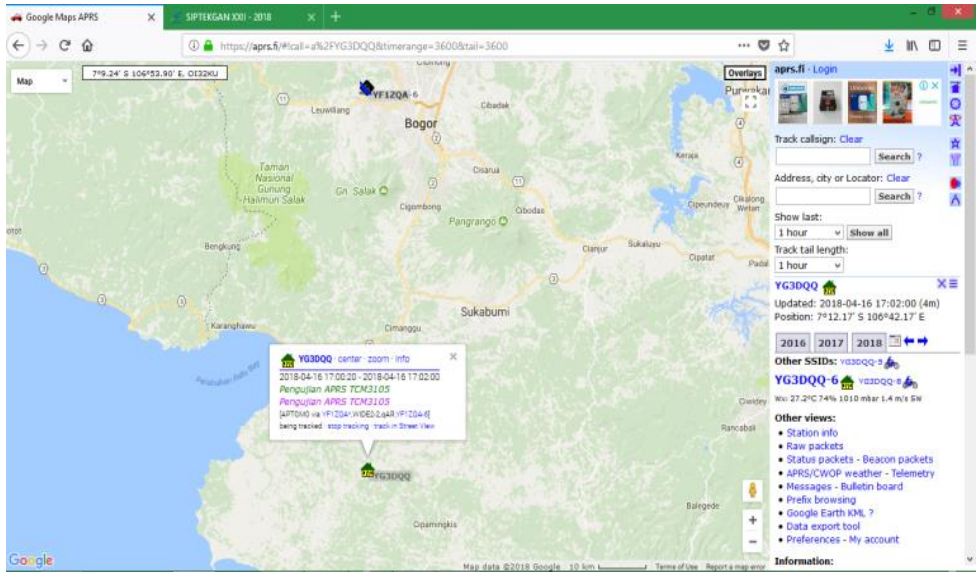

Figure 11. The APRS Packet Was Also Successfully Transmitted to The Aprs.fi
Through Internet Gateway (I-Gate) from YF1ZQA APRS Station